\documentclass[twoside,twocolumn,english,aps,manuscript,10pt,showpacs,amssymb,amsmath]{revtex4}
\usepackage[T1]{fontenc}
\usepackage[latin9]{inputenc}
\usepackage[letterpaper]{geometry}
\usepackage{babel}
\usepackage{units}
\usepackage{amsthm}
\usepackage{amsmath}
\usepackage{amssymb}
\usepackage{graphicx}
\usepackage{esint}
\usepackage[unicode=true,pdfusetitle,
 bookmarks=true,bookmarksnumbered=true,bookmarksopen=true,bookmarksopenlevel=1,
 breaklinks=false,pdfborder={0 0 0},backref=false,colorlinks=false]
 {hyperref}
\usepackage{breakurl}

\makeatletter
\@ifundefined{textcolor}{}
{%
 \definecolor{BLACK}{gray}{0}
 \definecolor{WHITE}{gray}{1}
 \definecolor{RED}{rgb}{1,0,0}
 \definecolor{GREEN}{rgb}{0,1,0}
 \definecolor{BLUE}{rgb}{0,0,1}
 \definecolor{CYAN}{cmyk}{1,0,0,0}
 \definecolor{MAGENTA}{cmyk}{0,1,0,0}
 \definecolor{YELLOW}{cmyk}{0,0,1,0}
}
\numberwithin{equation}{section}
\numberwithin{figure}{section}

\makeatother

\begin{document}

\title{\noindent {\LARGE Connection between maximum-work and maximum-power
thermal cycles}}

\author{\textbf{\textsl{\normalsize Julian Gonzalez-Ayala, L. A. Arias-Hernandez,
F. Angulo-Brown}}\textsl{}\\
\textsl{\normalsize Departamento de F\'isica, Escuela Superior de F\'isica
y Matem\'aticas, Instituto Polit\'ecnico Nacional, }\\
\textsl{\normalsize Edif. No. 9, U.P. Zacatenco, 07738, M\'exico D.
F., M\'exico }}
\begin{abstract}
We propose a new connection between maximum-power Curzon-Ahlborn thermal
cycles and maximum-work reversible cycles. This linkage is built through
a mapping between the exponents of a class of heat transfer laws and
the exponents of a family of heat capacities depending on temperature.
This connection leads to the recovery of known results and to a wide
and interesting set of new results for a class of thermal cycles.
Among other results we find that it is possible to use analytically
closed expressions for maximum-work efficiencies to calculate good
approaches to maximum-power efficiencies.
\end{abstract}


\maketitle
As it is well known, in 1975 \cite{C&A} Curzon-Ahlborn (CA) found
that an endoreversible Carnot-like thermal engine in which the isothermal
branches of the cycle are not in thermal equilibrium with their corresponding
heat reservoirs at absolute temperatures $T_{1}$ and $T_{2}<T_{1}$
has an efficiency at the maximum-power (MP) regime given by

\begin{equation}
\eta_{CA}=1-\sqrt{\frac{T_{2}}{T_{1}}}.\label{eq:C&A eff}
\end{equation}

This equation was obtained by using the so-called Newton law of cooling
to model the heat exchanges between the heat reservoirs and the working
fluid along the isothermal branches of the cycle. In fact, Eq.(\ref{eq:C&A eff})
was obtained previously by Novikov \cite{Novikov} in a context very
close to finite-time thermodynamics (FTT). Later, within the context
of FTT some authors \cite{Chen 1,Chen 2,P=0000E1ez-Angulo JAP2,Jhon Ross,De Vos}
demonstrated that Eq. (\ref{eq:C&A eff}) is only valid for the Newtonian
heat exchanges. As a matter of fact the CA-efficiency stems from taking
into account constant conductances and a linear dependence between
the heat fluxes and the working substance temperatures along the isothermal
branches of the cycle \cite{Chen 1,Chen 2,P=0000E1ez-Angulo JAP2,De Vos}.
Once the linearity in the heat transfer law is dropped, the square
root term in the MP-efficiency ($\eta_{MP}$) is lost. Very recently,
this fact has been widely confirmed by many authors working on the
first order irreversible thermodynamics \cite{Broeck 2005,Calvo et. al. con Luis,Calvo con Borja},
microscopic \cite{Seiffert stochastic microscopic} and mesoscopic
\cite{Z. Tu} heat engines. On the other hand, Eq.(\ref{eq:C&A eff})
also was obtained for some reversible thermal cycles performing at
maximum-work (MW) regime, such as the Otto and Joule-Brayton cycles
\cite{Harvey S. Leff. }. These coincident results for the CA, Otto
and Joule-Brayton cycles motivated Landsberg and Leff (LL) to propose
that the CA-efficiency possesses a nearly universal behavior for a
certain class of thermodynamic cycles operating at MW. This result
was achieved by means of a generalized cycle which reduces to the
Otto, Joule-Brayton, Diesel and some other known cycles \cite{Landsberg y Leff}.
Clearly, the kind of universality of the CA-efficiency claimed by
LL is not of the class of the true universality of Carnot efficiency,
$\eta_{C}$ \cite{Zemansky}. The square root term (SRT) observed
in the CA-efficiency can be found in other processes of energy conversion,
such as the so-called water powered machine, which mixes two steady
streams of hot and cold water to produce an output stream of warm
water at maximum kinetic energy \cite{LAPEN}. In fact, the role of
the SRT of temperatures is more general and it appears also in some
irreversible processes such as the irreversible cooling or heating
of an ideal gas initially at temperature $T_{i}$ in contact with
a series of auxiliary reservoirs to reach the final temperature $T_{f}$
of a main heat reservoir, the SRT appears when the generation of entropy
of this process is minimized \cite{Portugueses}. As it can be seen,
the SRT is found in several thermal processes (reversible or irreversible)
subject to some extremal conditions. A result less known is that corresponding
to the way the square root is lost in the case of reversible cycles
operating at MW regime. In 1989 \cite{Landsberg y Leff}, LL first
studied a cycle formed by two adiabatic processes and two paths with
constant heat capacities $C>0$ of the working fluid (see Fig. 1).
This reversible cycle operating under MW conditions has an efficiency
given by $\eta_{CA}=1-\sqrt{\tau}$, where $\tau=T_{-}/T_{+}$ is
the ratio between the minimum and maximum temperatures of the cycle
(see Fig. 1). Actually, the first author in finding this expression
for a MW engine was Chambadal \cite{Chambadal}. LL \cite{Landsberg y Leff}
generalized the model of Fig. 1, to encompass a family of symmetric
and asymmetric reversible cycles which have a MW efficiency that do
no deviate from $\eta_{CA}$ more than $14\%$. This behavior was
referred to as a near universality property of $\eta_{CA}$. However,
for the case of reversible cycles performing at MW, we will demonstrate
that the CA-efficiency is lost when constant heat capacities are not
used, in a similar way as it occurs in FTT, where the SRT in the CA-efficiency
is related only to constant conductances and to a linear law of heat
transfer.

We start with a heat capacity of the form $C=aT^{n}$ where $a$ is
a constant and $n$ a real number. Following the cycle depicted in
Fig. 1, we have after integrating the heat capacity over the temperature, 

\begin{equation}
Q_{in}=\begin{cases}
\begin{array}{c}
a\; ln\left(\frac{T_{+}}{T}\right)\\
\frac{aT_{+}^{n+1}}{n+1}\left[1-\left(\frac{T}{T_{+}}\right)^{n+1}\right]
\end{array} & \begin{array}{c}
n=-1\\
n\neq-1
\end{array}\end{cases},\label{eq:Qin}
\end{equation}

and

\begin{equation}
Q_{out}=\begin{cases}
\begin{array}{c}
b\; ln\left(\frac{T_{-}}{T'}\right)\\
\frac{bT_{-}^{n+1}}{n+1}\left[1-\left(\frac{T'}{T_{-}}\right)^{n+1}\right]
\end{array} & \begin{array}{c}
n=-1\\
n\neq-1
\end{array}\end{cases},\label{eq:Qout}
\end{equation}

where $b$ is also a constant and might be different from $a$. This
case represents a more general scenario for cyclic processes following
Fig. 1. The adjustable temperatures $T$ and $T'$are coupled because
the fluid\textquoteright{}s entropy change per reversible cycle is
zero, i.e.

\begin{equation}
\Delta S=\int_{T}^{T_{+}}\frac{Cd\mathbb{\mathbb{T}}}{\mathbb{T}}+\int_{T'}^{T_{-}}\frac{Cd\mathbb{T}}{\mathbb{T}}=0,\label{eq:DS}
\end{equation}

which leads to

\begin{equation}
T'=\begin{cases}
T_{+}\left(\frac{T_{-}}{T}\right)^{\gamma} & n=0\\
\left[T_{+}^{n}+\gamma\left(T_{-}^{n}-T^{n}\right)\right]^{\frac{1}{n}} & n\neq0
\end{cases},\label{eq:T'}
\end{equation}

where $\gamma=b/a$. Because the change in the total internal energy
is zero, the work done per cycle $W=|Q_{in}|-|Q_{out}|$ satisfies

\begin{equation}
W=\begin{cases}
aT_{+}\left[1-\left(\frac{T_{-}}{T}\right)^{\gamma}\right]+b\left(T_{-}-T\right) & n=0\\
\begin{array}{c}
a\; ln\left[1+\gamma T_{+}\left(T_{-}^{-1}-T^{-1}\right)\right]+\\
b\; ln\left(\frac{T_{-}}{T}\right)
\end{array} & \text{ \ensuremath{n}}=-1\\
\begin{array}{c}
\frac{aT_{+}^{n+1}}{n+1}\left\{ 1-\left[1+\gamma\left(\frac{T_{-}^{n}}{T_{+}^{n}}-\frac{T^{n}}{T_{+}^{n}}\right)\right]^{\frac{n+1}{n}}\right\} \\
+\frac{bT_{-}^{n+1}}{n+1}\left(1-\frac{T^{n+1}}{T_{-}^{n+1}}\right)
\end{array} & n\neq0,-1
\end{cases}\label{eq:W cero entropia}
\end{equation}

\begin{figure}[t]
\noindent \begin{centering}
\includegraphics[width=4cm]{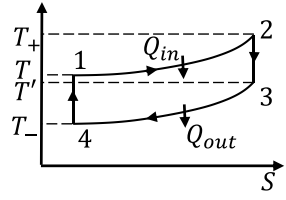}
\par\end{centering}

\caption{$T-S$ diagram of a cycle formed with two adiabats and two other processes
with $C>0$. }
\end{figure}

By maximizing $W$ with respect to $T$ we find that $T^{*}$ and
$T'^{*}$ are given by

\begin{equation}
T^{*}=T'^{*}=\begin{cases}
\left(T_{+}T_{-}^{\gamma}\right)^{\frac{1}{1+\gamma}} & n=0\\
\frac{\left(1+\gamma\right)T_{+}T_{-}}{\gamma T_{+}+T_{-}} & \text{ \ensuremath{n=-1}}\\
\left(\frac{T_{+}^{n}+\gamma T_{-}^{n}}{1+\gamma}\right)^{\frac{1}{n}} & n\neq0,-1
\end{cases}.\label{eq:T*_T1T3}
\end{equation}

Then, as a result, the named symmetric cycles ($\gamma=1$) and asymmetric
cycles ($\gamma\neq1$) fulfill that $T^{*}=T'^{*}$. Therefore, from
those expressions we immediately find the efficiency $\eta=W/Q_{in}$
under MW conditions; that is,

\begin{equation}
\eta_{MW}=\begin{cases}
1-\gamma\tau^{\frac{\gamma}{1+\gamma}}\left(\frac{1-\tau^{\frac{1}{1+\gamma}}}{1-\tau^{\frac{\gamma}{1+\gamma}}}\right) & n=0\\
1+\gamma\frac{ln\left(\frac{\tau+\gamma}{1+\gamma}\right)}{ln\left(\frac{1+\frac{\gamma}{\tau}}{1+\gamma}\right)} & \text{ \ensuremath{n=-1}}\\
1-\gamma\tau^{n+1}\frac{\left(\frac{\tau^{-n}+\gamma}{1+\gamma}\right)^{\frac{n+1}{n}}-1}{1-\left(\frac{1+\gamma\tau^{n}}{1+\gamma}\right)^{\frac{n+1}{n}}} & n\neq0,-1
\end{cases}\label{eq:eta max}
\end{equation}

Clearly, Eq. (\ref{eq:eta max}) only reproduces the $\eta_{CA}$-efficiency
for a few combinations of $\gamma$ and $n$. In Fig. 2 we depict
the plot of $\eta_{MW}$ versus the exponent $n$ considering $\gamma=1$
(symmetric scenario). In this figure we observe that $\eta_{CA}$
is obtained only for two values of $n$, the known case $n=0$ (constant
heat capacity) and also at $n=-1/2$, which is a novel case with MW-efficiency
given by $\eta_{CA}$. Another relevant case is that with $n=-1/4$,
where the MW-efficiency takes its maximum value.

\begin{figure}[t]
\noindent \begin{centering}
\includegraphics[bb=0bp 0bp 256bp 96bp,scale=0.9]{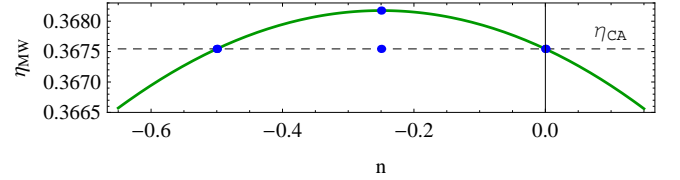}
\par\end{centering}

\caption{$\eta_{MW}$ vs $n$ with $\gamma=1$ and $\tau=2/5$. The cases $n=0,-1/2$
reproduce the well-known CA efficiency. Notice that there is a maximum
at $n=-1/4$.}
\end{figure}

By fixing the value of $\tau$ it is possible to find out (Fig. 3)
that the CA-efficiency does not have a special behavior with respect
to other efficiencies, in the sense that it does not represent the
maximum value for an efficiency at MW. In Fig. 3, we can observe how
the case $n=0$ only reaches the value $\eta_{CA}$ for $\gamma=1$;
that is, the case of symmetric cycles with constant heat capacities
performing at MW \cite{Harvey S. Leff. ,Landsberg y Leff}. Besides,
in this figure we can see how the $\eta_{CA}$ value for MW-efficiency
is also obtained for asymmetric cases ($\gamma\neq1$) and $n\neq0$.
It is quite remarkable that the unique case independent of $\gamma$
is that with $n=-1/2$; that is $\eta_{MW}\left(n=-1/2\right)=\eta_{CA}$
for any value of $\gamma$. In any case, for a cycle as shown in Fig.
1 with $C=aT^{-\nicefrac{1}{2}}$ in the process $1\rightarrow2$
and $C=bT^{-1/2}$ in the process $3\rightarrow4$, the true cycle
is characterized by the $\eta_{CA}$ at MW for any value of $\gamma$.
However, within the context of FTT, the cycle with MP-efficiency given
by $\eta_{CA}$ and independent of $\gamma'=\beta/\alpha$ (being
$\alpha$ and $\beta$ heat conductances, see Eqs. (\ref{eq:Q-FTT})
and (\ref{eq:Q-FTTout})) is indeed the Curzon and Ahlborn cycle. 

\begin{figure}[t]
\noindent \begin{centering}
\includegraphics[bb=10bp 10bp 240bp 95bp,scale=0.9]{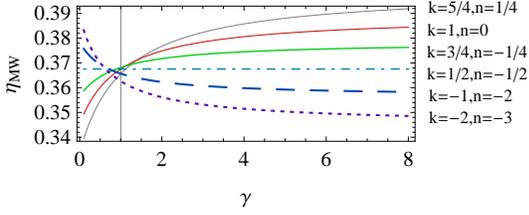}
\par\end{centering}

\caption{$\eta_{MW}$ vs $\gamma$ with $\tau=2/5$. The case $n=0$ reproduces
the CA efficiency only when $\gamma=1$ as it is known, but the case
$n=-1/2$ reproduces the same result for any $\gamma$.}
\end{figure}

\begin{figure}[t]
\noindent \begin{centering}
\includegraphics[scale=0.9]{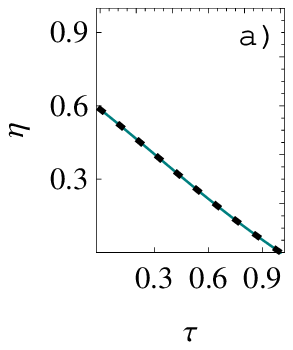}\includegraphics[scale=0.9]{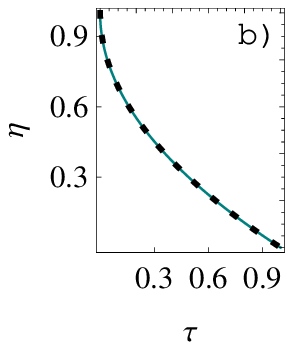}\includegraphics[scale=0.9]{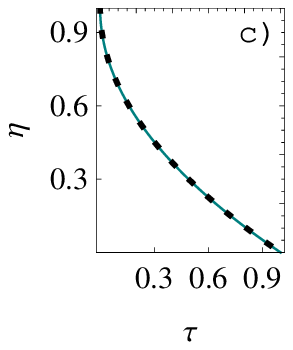}
\par\end{centering}

\caption{Comparison between the MW (thick-line) and their corresponding MP
(dotted-line) efficiencies. The depicted efficiencies are for symmetric
cases $\gamma=1$: a).- $n=-2$ in Eqs. (\ref{eq:Qin}) and (\ref{eq:Qout})
and $k=-1$ in Eqs. (\ref{eq:Q-FTT}) and (\ref{eq:Q-FTTout}); b).-
$n=-1/2$ and $k=1/2$ ; c).- $n=0$ and $k=1$. }
\end{figure}

Through their papers on reversible cycles performing at MW, LL \cite{Harvey S. Leff. ,Landsberg y Leff}
established a bridge with FTT-MP-cycles, basically by means of the
CA-efficiency. In recent years several authors \cite{Broeck 2005,Calvo et. al. con Luis,Calvo con Borja,Broeck  particle transport,Norma Optimization criteria,Wang Eff max power,Z. Tu con Wang,Esposito 2010 Low Diss}
have renewed the interest on the CA-efficiency. The discussion on
this famous formula has been mainly turned on to its possible universal
nature within the context of finite-time cycles. Practically since
the beginning of FTT as an active discipline, the limited validity
of the CA-efficiency was established \cite{Chen 1,Chen 2,P=0000E1ez-Angulo JAP2,Jhon Ross,De Vos}.
Nonetheless, at the end of the 1980's the CA-efficiency was found
linked to MW reversible cycles \cite{Harvey S. Leff. ,Landsberg y Leff}
and recently with its role in microscopic \cite{Seiffert stochastic microscopic},
mesoscopic \cite{Z. Tu} and macroscopic thermal cycles \cite{Broeck 2005,Calvo et. al. con Luis},
and therefore the CA-efficiency has gained new insights, that have
to do with a possible universality at least at the first few terms
in the Taylor expansion of the efficiency as a function of $\eta_{C}=1-\tau$
\cite{Z. Tu,Broeck  particle transport,Norma Optimization criteria,Wang Eff max power}.
Clearly, Eq. (\ref{eq:eta max}) admits this treatment, leading to
a series in terms of $\eta_{C}$ as functions of $\gamma$ and $n$
at any level of approximation. The series for the efficiencies in
Eq (\ref{eq:eta max}) are given by the following expressions,

\noindent \begin{center}
\begin{equation}
\eta^{*}=\begin{cases}
\frac{\eta_{C}}{2}+\frac{\left(1+2\gamma\right)\eta_{C}^{2}}{12\left(1+\gamma\right)}+\frac{\left(1+2\gamma\right)\eta_{C}^{3}}{24\left(1+\gamma\right)}+\mathcal{O}[\eta_{C}]^{4} & n=0\\
\frac{\eta_{C}}{2}+\frac{\left(2+\gamma\right)\eta_{C}^{2}}{12\left(1+\gamma\right)}+\frac{\left(2+2\gamma+\gamma^{2}\right)\eta_{C}^{3}}{24\left(1+\gamma\right)^{2}}+\mathcal{O}[\eta_{C}]^{4} & \text{ \ensuremath{n=-1}}\\
\begin{array}{c}
\frac{\eta_{C}}{2}+\frac{\left(1-n+2\gamma+n\gamma\right)\eta_{C}^{2}}{12\left(1+\gamma\right)}+\qquad\qquad\qquad\quad\\
\frac{\left(1-n+3\gamma-n\gamma-2n^{2}\gamma+2\gamma^{2}+n\gamma^{2}\right)\eta_{C}^{3}}{24\left(1+\gamma\right)^{2}}+\mathcal{O}[\eta_{C}]^{4}
\end{array} & n\neq0,-1
\end{cases},\label{eq:eta (etaCarnot)}
\end{equation}

\par\end{center}

The linear term is really the same in every case, that strengthen
the idea that this is a characteristic for cycles operating in the
maximum work regime. When we restrict $\gamma$ to the symmetric case,
$\gamma=1$, then the efficiencies are

\noindent \begin{center}
\begin{equation}
\eta^{*}=\begin{cases}
\frac{\eta_{C}}{2}+\frac{\eta_{C}^{2}}{8}+\frac{\eta_{C}^{3}}{16}+\mathcal{O}[\eta_{C}]^{4} & n=0\\
\frac{\eta_{C}}{2}+\frac{\eta_{C}^{2}}{8}+\frac{5\eta_{C}^{3}}{96}+\mathcal{O}[\eta_{C}]^{4} & \text{ \ensuremath{n=-1}}\\
\frac{\eta_{C}}{2}+\frac{\eta_{C}^{2}}{8}+\frac{\left(6-n-2n^{2}\right)\eta_{C}^{3}}{96}+\mathcal{O}[\eta_{C}]^{4} & n\neq0,-1
\end{cases},\label{eq:eta (etaCarnot)-1}
\end{equation}

\par\end{center}

In this case the coincidence extends up to quadratic order in $\eta_{C}$,
such as it occurs for MP strong coupling models that possess a left-right
symmetry \cite{Broeck  particle transport,Broeck universality}.

Following the spirit of the articles of LL \cite{Harvey S. Leff. ,Landsberg y Leff},
we may wonder: Could it be possible to link the results obtained with
heat capacities depending on temperature given by Eq. (8) with finite-time
cycles of the CA-type? We suggest that this connection is indeed possible.
As it is well known, in reversible cycles of the Otto and Joule-Brayton
type, $Q_{in}$ and $Q_{out}$ correspond to quasi-static processes
accomplished by means of an infinite set of auxiliary heat reservoirs
that lead the working substance temperature from $T$ to $T_{+}$
and from $T'$ to $T_{-}$, respectively. These heat quantities are
calculated by means of integrals that lead to Eqs. (\ref{eq:Qin})
and (\ref{eq:Qout}). In the case of FTT-cycles as the CA-engine,
$Q_{in/out}$ per cycle are given by irreversible heat transfer laws
depending on the conductances ($\alpha$, $\beta$) and the temperatures
of the corresponding heat reservoirs and the working substance. As
asserted by Wang and Tu \cite{Z. Tu con Wang}, for the CA-cycle,
along both ``isothermal'' branches, the effective temperature of
the working substance can vary. Thus, we can propose that in a $T-S$
plane, the CA-cycle follows a topologically equivalent diagram as
that of Fig. 1. For FTT-cycles, heat transfer laws of the form

\begin{equation}
\dot{Q}_{in}=\alpha T_{1}^{k}\left[1-\left(\frac{T_{1W}}{T_{1}}\right)^{k}\right],\label{eq:Q-FTT}
\end{equation}

\begin{equation}
\dot{Q}_{out}=\beta T_{2}^{k}\left[\left(\frac{T_{2W}}{T_{2}}\right)^{k}-1\right],\label{eq:Q-FTTout}
\end{equation}

are typically used, where $T_{1W/2W}$ are the working substance temperatures,
$T_{1/2}$ are the heat reservoir temperatures and $k$ is a real
number. Although evidently the conceptual meaning of the heat quantities
($Q_{in/out}$) is different within the framework of reversible cycles
and FTT cycles respectively, it is quite remarkable how their corresponding
efficiencies have a good agreement for not arbitrary couples of $n$
and $k$ values. As is well known, $\dot{Q}_{in/out}$, power output
and MP-efficiency for CA-engines with heat transfer laws given by
Eqs. (\ref{eq:Q-FTT}) and (\ref{eq:Q-FTTout}), are numerically calculated
in an easy and direct manner by maximizing the power output $P$ given
by the following equation\cite{JAP Angulo Arias,Open Systems,Angulo Arias Rev Mex 94},

\begin{equation}
P\left(\eta\right)=\frac{\alpha T_{1}\gamma\eta\left[\left(1-\eta\right)^{k}-\tau^{k}\right]}{1-\eta+\gamma\left(1-\eta\right)^{k}}.\label{eq:Potencia}
\end{equation}

with respect to the efficiency $\eta$. In Fig. 4, we can observe
the excellent fitting between three MW-efficiencies and their corresponding
FTT-MP-cases. Clearly, the mapping between $n$ and $k$ is given
by $k\rightarrow n+1$, which stems from associating the exponents
of Eqs. (\ref{eq:Qin}) and (\ref{eq:Qout}) with those of Eqs. (\ref{eq:Q-FTT})
and (\ref{eq:Q-FTTout}).

\begin{figure*}[t]
\begin{centering}
\includegraphics[scale=0.9]{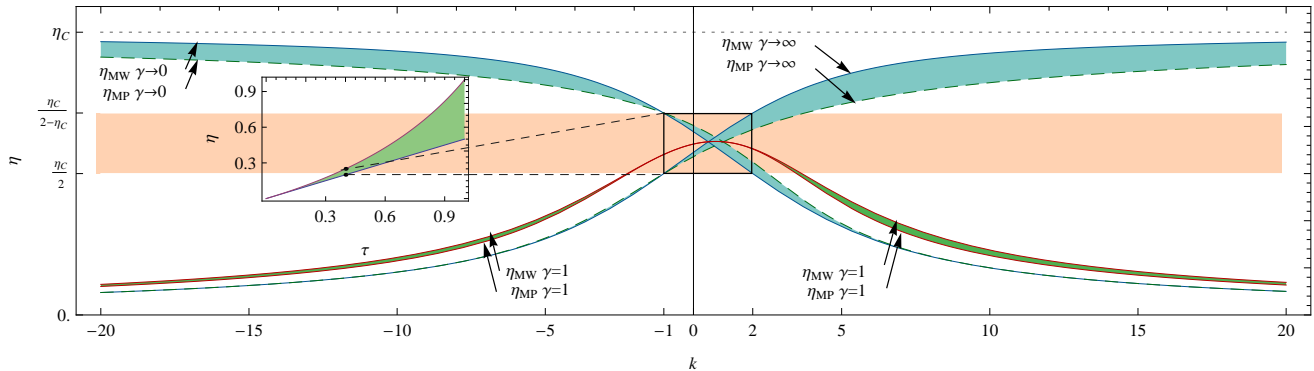}
\par\end{centering}

\caption{The bounds for the $\eta_{MW}$ and $\eta_{MP}$ are given by the
$\chi$-shaped curves for $\gamma\rightarrow0$ and $\gamma\rightarrow\infty$,
with $\tau=2/5$. For $k=-1$ ($n=-2$) the well-known limits reported
by Esposito et. al. \cite{Esposito 2010 Low Diss} are reproduced
for both $\eta_{MW}$ and $\eta_{MP}$ (see inset). These limits are
also reproduced by $\eta_{MW}$ at $n=1$. For the symmetric cases
($\gamma=1$) $\eta_{MW}\geq\eta_{MP}$ with a maximum at $k=3/4$
($n=-1/4$, see also Fig. 2). Over the $\chi$-shaped curves we can
observe some regions where $\eta_{MW}\geq\eta_{MP}$. Notice that
the upper asymptote is the value of $\eta_{C}$ and the lower one
is zero. We used numerical calculations for $\eta_{MP}$. }
\end{figure*}

The fitting between $\eta_{MW}$ and $\eta_{MP}$ goes from excellent
(symmetric cases) to very good (asymmetric cases) (see Figs. 4 and
5), in such a way that the analytical closed expressions of $\eta_{MW}$
given by Eq. (\ref{eq:eta max}) can be used as reliable first approximations
for the FTT corresponding cycles, which commonly have to be calculated
by means of numerical methods. 

Another remarkable fact is that the MW-efficiency when $n=-2$ and
the corresponding MP-efficiency at $k=-1$ are exactly the same for
any value of $\gamma$ (from now on $\gamma=\gamma'$), that is, 

\begin{equation}
\eta_{MW}\left(n=-2\right)=\eta_{MP}\left(k=-1\right)=\frac{\eta_{C}}{2-\left(1-\sqrt{\frac{\gamma+\tau^{2}}{1+\gamma}}\right)}.\label{eq:eta n=00003D-2=00003D=00003D eta k=00003D-1}
\end{equation}

Which has two limits: $\gamma\rightarrow0$ and $\gamma\rightarrow\infty$
bounding the possible values of the efficiency for a given $\tau$,
at $n=-2$ ($k=-1$)
\begin{equation}
\underset{\gamma\rightarrow\infty}{Lim}\left(\eta_{MW}\right)=\frac{\eta_{C}}{2}<\eta_{MW/MP}<\frac{\eta_{C}}{2-\eta_{C}}=\underset{\gamma\rightarrow0}{Lim}\left(\eta_{MW}\right).\label{eq:limits eta}
\end{equation}

Recently, some authors have underscored the importance of these limits
(first found in \cite{Chen 1}), which have been reported within different
contexts, such as, a stochastic heat engine \cite{Seiffert stochastic microscopic},
a low-dissipation Carnot engine \cite{Esposito 2010 Low Diss} and
for a linear irreversible Carnot-like heat engine \cite{Z. Tu con Wang}.
However, we shall see below that these limits are only of a particular
validity among a numerous set of limits for different values of $k$
(or $n$). On the basis of Eq. (\ref{eq:eta max}) we obtain the limits
of $\eta_{MW}$ for $\gamma\rightarrow0$ and $\gamma\rightarrow\infty$
which bound the values of $\eta_{MW}$. These $\chi$-shaped curves
(continuous curves) are depicted in Fig. 5 where the corresponding
$\eta_{MP}$ curves (large dashed) are also showed along with the
symmetric cases for both efficiencies. We have used well-known FTT
numerical methods to plot the $\eta_{MP}$ curves \cite{Angulo Arias Rev Mex 94,Open Systems,JAP Angulo Arias}.

Some interesting facts can be remarked from this figure: as mentioned
before, for $n=-2$ ($k=-1$) both efficiencies $\eta_{MW}$ and $\eta_{MP}$
have the same limits when $\gamma\rightarrow0$ and $\gamma\rightarrow\infty$.
Interestingly, the $\chi$-shaped curve corresponding to the $\eta_{MW}$
efficiencies have an exact specular symmetry with respect to the value
$n=-1/2$ ($k=1/2$). At this point, the upper and lower limits of
the efficiency are the same because $\eta_{MW}\left(n=-1/2\right)$
does not depend on $\gamma$. This specular symmetry has as a consequence
that both limits given by Eq. (\ref{eq:limits eta}) are also found
in the MW-efficiency for $n=1$ ($k=2$, see Fig. 5). On the other
hand, the $\chi$-shaped curve for the $\eta_{MP}$ efficiencies does
not have a specular symmetry with respect to the crossing point at
$k=1$, where both limits ($\gamma\rightarrow0$ and $\gamma\rightarrow\infty$)
are the same because for $k=1$ $\eta_{MP}=\eta_{CA}$ is independent
of $\gamma$. This lack of symmetry precludes that the limits given
by Eq. (\ref{eq:limits eta}) appear for any other $k\neq-1$. It
should be noted that at the left side and at the right side of the
crossing points over the MW and MP $\chi$-shaped curves, the lower
and upper limits are interchanged. For the specular symmetric MW case,
when exchanging $n\rightarrow-\left(n+1\right)$ ($k\rightarrow-(k-1)$),
both limits have the same value, but they are inverted ($\gamma\rightarrow0$
is replaced by $\gamma\rightarrow\infty$ and vice versa). There is
another fact of great interest about the $\chi-$shaped curves. For
heat transfer laws with approximately $k\in\left(-10,10\right)$,
we can observe that in some regions $\eta_{MW}<\eta_{MP}$. This inequality
is not an artifact of numerical solutions for $\eta_{MP}$, because
in that region exist some cases where the inequality is an exact analytical
result, for example, for $k=1/2$ ($n=-1/2$). However, clearly, if
in addition to the heat fluxes many other irreversibilities are considered,
the above mentioned inequality should be inverted. For the Stefan-Boltzmann
case, that is, $k=4$ ($n=3$); the upper and lower limits, both for
$\eta_{MW}$ and $\eta_{MP}$ are not the limits given by Eq. (\ref{eq:limits eta})
and additionally they are inverted, in such a way that for $\gamma'\rightarrow\infty$
a  M\"user--type engine ($\beta\gg\alpha$) is obtained \cite{De Vos Solar Energy Conversion}
and $\eta_{MW}>\eta_{MP}$.

In Fig. 5 a large range of values of $k$ (or $n$) is considered,
showing that at the same limits $\gamma\rightarrow0$, $\gamma\rightarrow\infty$
and $\gamma\rightarrow1$ the values of the MP and MW--efficiencies
are very close to each other, strengthening the idea that the analytical
forms of the MW-efficiencies are a good approximation to the corresponding
MP efficiencies. In addition, for all symmetric cases $\eta_{MW}\geq\eta_{MP}$
and both convex curves ($\gamma=1$) tend to zero when $|n|,|k|\rightarrow\infty$.
It is noteworthy that the superior branches of the $\chi-shaped$
curves for both $\eta_{MW}$ and $\eta_{MP}$ tend asymptotically
to $\eta_{C}$ and both inferior branches tend to zero when $|n|,|k|\rightarrow\infty$.
These results are consequence of Eq. (\ref{eq:eta max}) and the numerical
solutions of FTT--MP--efficiencies. The physical consistency of all
these asymptotic limits stems from the restrictions imposed by Eqs.
(\ref{eq:T'}) and (\ref{eq:T*_T1T3}) and the first and second laws
of thermodynamics.

In summary, in this article we have demonstrated that there exists
a strong and rich relationship between a class of reversible MW--cycles
and finite--time MP--cycles of the CA--type. This connection opens
a wide spectrum of interesting results containing known facts and
some new findings. All of our results suggest that behind the remarkable
agreement of the mentioned connection there is a kind of extended
endoreversibility contained in Figs. (1) and (5) for very low-dissipation
cycles, as those of very large compression ratios \cite{Jhon Ross,Esposito 2010 Low Diss}.
\begin{acknowledgments}
We want to thank partial support from COFAA-SIP-EDI-IPN and SNI-CONACYT,
M\'EXICO.\end{acknowledgments}

\end{document}